\documentclass[aps,pra,floatfix,superscriptaddress,twocolumn,letterpaper]{revtex4}

\usepackage[latin1]{inputenc}
\usepackage[english]{babel}
\usepackage{graphicx, amssymb, amsmath, times, siunitx, gensymb, xcolor, multirow, rotating}
\usepackage[stretch=10]{microtype}

\newcommand{\tr}{{\rm Tr\thinspace}}
\newcommand{\bra}[1]{\left\langle{#1}\right\vert}
\newcommand{\ket}[1]{\left\vert{#1}\right\rangle}

\newcommand{\dg}{^\dagger}

\newcommand{\etal}{\textit{et al.}~}

\newcommand{\BQIC}{Berkeley Center for Quantum Information and Computation, Berkeley, California 94720 USA}
\newcommand{\DeptPhys}{Department of Physics, University of California, Berkeley, California 94720 USA}
\newcommand{\DeptChem}{Department of Chemistry, University of California, Berkeley, California 94720 USA}
\newcommand{\Sandia}{Department of Scalable and Secure Systems Research, Sandia National Laboratories, Livermore, CA 94550 USA}
\newcommand{\UlmPlenio}{Institut für Theoretische Physik, Universität Ulm, Albert-Einstein-Allee 11, D-89069 Ulm, Germany}
\newcommand{\UlmCalarco}{Institut für Quanteninformationsverarbeitung, Universität Ulm, Albert-Einstein-Allee 11, D-89069 Ulm, Germany}
\newcommand{\LENS}{LENS and Dipartimento di Fisica e Astronomia, Università di Firenze, I-50019 Sesto Fiorentino, Italy}
\newcommand{\QSTAR}{QSTAR, Largo Enrico Fermi 2, I-50125 Firenze, Italy}

\begin{document}

\title{Realistic and verifiable coherent control of excitonic states in a light harvesting complex}

\author{Stephan Hoyer}
% \author{S. Hoyer}
\affiliation{\BQIC}
\affiliation{\DeptPhys}

\author{Filippo Caruso}
% \author{F. Caruso}
\affiliation{\UlmPlenio}
\affiliation{\LENS}
\affiliation{\QSTAR}

\author{Simone Montangero}
% \author{S. Montangero}
\affiliation{\UlmCalarco}

\author{Mohan~Sarovar}
% \author{M. Sarovar}
\affiliation{\Sandia}

\author{Tommaso Calarco}
% \author{T. Calarco}
\affiliation{\UlmCalarco}

\author{Martin B. Plenio}
% \author{M.B. Plenio}
\affiliation{\UlmPlenio}

\author{K. Birgitta Whaley}
% \author{K.B. Whaley}
\affiliation{\BQIC}
\affiliation{\DeptChem}

\date{\today}

\begin{abstract}
We explore the feasibility of coherent control of excitonic dynamics in light harvesting complexes,
analyzing the limits imposed by the open nature of these quantum systems. We establish feasible targets for phase and phase/amplitude control of the electronically excited state populations in the Fenna-Mathews-Olson (FMO) complex and analyze the robustness of this control
with respect to orientational and energetic disorder, as well as decoherence arising from coupling to the protein environment. We further present two possible routes to verification of the control target, with simulations for the FMO complex showing that steering of the excited state is experimentally verifiable either by extending excitonic coherence or by producing novel states in a pump-probe setup. Our results provide a first step toward coherent control of these
complex biological quantum systems in an ultrafast spectroscopy setup.
\end{abstract}
\maketitle

\section{Introduction}
\label{sec:introduction}
The control of atomic and molecular processes using coherent sources of radiation is a well established experimental technique. Particularly successful have been implementations that aim to control the non-equilibrium dynamics of highly coherent quantum systems, e.g. internal and external states of cold atoms \cite{Chu:2002wi}, and ones that aim to control the equilibrium states resulting from controlled mixed (coherent and incoherent) dynamics such as those dictating chemical reactions \cite{Shapiro:1997tm}.
On the other hand, few experiments have attempted to control the non-equilibrium states of systems undergoing mixed dynamics.
Indeed, recent theoretical work in the field of quantum control has actively investigated questions related to such \textit{open quantum systems} (e.g. \cite{Brif:2010tc,Schmidt:2011bv,Dirr:2009}). Formulating optimal control protocols can be challenging for such systems.  In fact, it is usually difficult to even decide whether or not an open quantum system is controllable (i.e., whether all states are reachable from an arbitrary initial state), given a model of its dynamics and control \cite{Dirr:2009}. Theoretical treatment of such systems is complicated by the fact that one cannot exploit a clear separation of timescales to restrict the dynamical model to purely coherent or incoherent dynamics. Instead one must incorporate both coherent dynamics and decoherence processes in a unified manner, e.g., using quantum master equation models.

In this paper we examine a paradigmatic open quantum system, electronic excited states in photosynthetic light harvesting complexes (LHCs) \cite{Bla-2002}, and investigate strategies for controlling the non-equilibrium dynamics of these excited states. Photosynthetic light harvesting complexes are typically composed of dense arrangements of pigment molecules, such as chlorophyll, embedded within protein backbones. Electronic excited states result from the absorption of photons by pigments in such pigment-protein structures. These states -- termed \textit{excitons} -- can be either localized on single pigments or delocalized across multiple pigments due to the strong electronic coupling between pigments. The excitation energy carried by these states is funneled to regions of the light harvesting complex that can initiate the decomposition of such excitons into separated free charge carriers. This energy transfer process, which is dictated by the dynamics of excitons, is extremely complex and has recently been shown to have significant quantum coherent character \cite{Eng.Cal.etal-2007,Col.Won.etal-2010,Pan.Hay.etal-2010}. Subsequent modeling and theoretical study \cite{Mohseni2008,Plenio2008,Ish.Fle-2009b,Reb.Moh.etal-2009a,Car.Chi.etal-2009,Sar.Ish.etal-2009,Rey:2013ih,Chin:2013ia,Caruso2010,Chin2010,Caruso:2010tm} have determined that the energy transfer process is described by a finely tuned balance of coherent and incoherent dynamical processes. Due to this partially coherent nature of the excitonic dynamics, the control of the energy transfer process in LHCs using laser fields is expected to be sensitive to both the temporal and spatial phase coherence of the controlling laser fields.
For these reasons, we regard the control of energy transfer in LHCs as a paradigmatic example of coherent control of mixed quantum dynamics in open quantum system dynamics.  Achieving control of excitonic dynamics in photosynthetic systems is a first step towards active control of energy transfer dynamics in complex organic molecular assemblies, a potentially important capability for artificial light harvesting technologies \cite{Sar:2013}. The ability to control excitonic dynamics in LHCs is also a potentially useful tool in the quest to develop a complete understanding of energy transfer in these complex systems.

Several previous studies have already addressed the coherent control of excitonic dynamics in LHCs. Herek \etal performed an early experiment demonstrating moderate control over energy transfer pathways in the LH2 light harvesting complex using shaped femtosecond laser pulses \cite{Her.Woh.etal-2002}. Theoretical modeling of this same LHC and calculation of optimal control fields to achieve enhanced fluorescence was performed in Ref.~\cite{Lami:2004ba}. These studies demonstrated coherent control of chemical products of light harvesting system, but not control of femtosecond electronic dynamics themselves. In contrast,
Brüggemann, May and co-workers performed a series of theoretical studies focusing on the formulation of optimal femtosecond pulses to control the excitonic dynamics of the Fenna Matthews Olson (FMO) complex \cite{Bruggemann:2004fi,Bruggemann:2006ih,Bruggemann:2007hv}. These studies aimed to localize excitation energy in regions of the FMO complex using shaped pulses with and without polarization control. Recently, Caruso \etal \cite{Caruso:2012hv} performed a theoretical study that focused on preparing various localized and propagating excitonic states of the FMO complex using shaped femtosecond pulses determined with the recently introduced CRAB optimization algorithm \cite{Doria:2011vm}.

In this work, we extend the studies of optimal control of excitonic dynamics in FMO by systematically analyzing the limitations to achievable control that are imposed by practical constraints for bulk, small ensemble and single complex experiments.  As an important complement, we also identify schemes for authentication of any such coherent control of excitonic dynamics by prediction of the signatures of coherent control in pump-probe spectra.
We close with a discussion of the outlook for further development of coherent control and its applications for analysis of excitonic coherence in biological systems.

\section{Model for laser excitation of FMO}
\label{sec:laser_FMO}

To model light-harvesting in photosynthetic systems, we use the Heitler-London model Hamiltonian \cite{Agrano1982,May2011}, written as a sum of terms including the electronic system (S), the vibrational reservoir (R) and light (L):
\begin{align}
	H &= H_\text{S} + H_\text{S-R} + H_\text{R} + H_\text{S-L}.
	\label{eq:H_full}
\end{align}
We use a Frenkel exciton model for the system,
\begin{align}
	H_\text{S} &= \sum_{n} \mathcal{E}_n a_n^\dagger a_n + \sum_{n\neq m} J_{nm} a_n^\dagger a_m
	\label{eq:H_S}
\end{align}
where $a_n$ is the annihilation operator for an electronic excitation on pigment $n$, $\mathcal{E}_n$ the excitation energy on pigment $n$ and $J_{nm}$ the dipole-dipole coupling between pigments $n$ and $m$. We treat these excitations as hard-core bosons (that is, not allowing double excitations of a single pigment), and restrict our consideration to only one possible excitation per pigment molecule. The reservoir is modeled as a collection of harmonic oscillators at thermal equilibrium,
\begin{align}
	H_\text{R} &= \sum_{\xi} \hbar \omega_\xi b_\xi^\dagger b_\xi
	\label{eq:H_R}
\end{align}
where $b_\xi$ denotes a reservoir annihilation operator and $\hbar\omega_\xi$ the energy of an excitation in reservoir mode $\xi$.
The system-reservoir coupling is then given by
\begin{align}
	H_\text{S-R} &= \sum_{n, \xi} \hbar \omega_\xi g_{n \xi} a_n^\dagger a_n (b_\xi + b_\xi^\dagger),
	\label{eq:H_S-R}
\end{align}
where $g_{n\xi}$ denotes the unit-less strength of the coupling between electronic excitation $n$ and bath mode $\xi$. All information regarding the reservoir is contained in the spectral density function $\mathcal{J}_n(\omega) = \sum_{\xi} g_{n\xi}^2 \delta(\omega - \omega_\xi)$.
Lastly, the system-light interaction is given by
\begin{align}
	H_\text{S-L} &= \vec{\mu} \cdot \vec{E}(t)
	\label{eq:H_S-L}
\end{align}
where
$\vec{E}(t)$ is the semi-classical electric field of the incident light and $\vec\mu = \sum_n \vec d_n (a_n + a_n\dg)$ is the transition dipole operator, where $\vec d_n$ is the transition dipole vector of pigment $n$.

We focus our investigations on the FMO complex of green sulfur bacteria, an extensively studied protein-pigment complex. Biologically, a monomer of the FMO complex serves as a ``quantum wire'' with seven chlorophyll pigments that transfers excitation energy from the chlorosome antenna towards the reaction center in a partially coherent manner \cite{Eng.Cal.etal-2007,Pan.Hay.etal-2010}. The crystal structure of the FMO protein is known, and the system has been subject to a large number of linear and non-linear spectroscopic measurements \cite{Milder2010}. Accordingly, the Hamiltonian of the system is quite well established.  Here we use the electronic Hamiltonian determined by Adolphs and Renger that includes Gaussian static (ensemble) disorder of full-width-at-half-maximum \SI{100}{cm^{-1}} for each transition energy $\mathcal{E}_n$ \cite{Adolphs2006}.
We model the system-reservoir coupling
by assuming that each pigment is coupled to an independent reservoir with identical Debye spectral densities $\mathcal{J}_n(\omega) = \mathcal{J}(\omega)$ of the form $\omega^2 \mathcal{J}(\omega) = \frac{2 \lambda \gamma \omega}{\omega^2 + \gamma^2} \Theta(\omega)$, where $\Theta$ denotes the Heaviside function and with reorganization energy $\lambda=\SI{35}{cm^{-1}}$ and bath relaxation rate $\gamma=1/(\SI{50}{fs})$
tuned to experimental values as in Ref. \cite{Ish.Fle-2009b}.

Because for FMO all of the key energy scales in the problem coincide ($\hbar \lambda \sim \hbar \gamma \sim J_{nm}$)
the Hamiltonian given by Eqs.~\eqref{eq:H_full}-\eqref{eq:H_S-L} does not fall in the range of validity of the standard perturbative approaches of either Förster or Redfield theories \cite{Ish.Fle-2009b}. This has made photosynthetic systems, and in particular the FMO complex, prototypes for alternative methods to solve open quantum systems \cite{Ishizaki:2009,ritschel2011efficient,huo2010iterative,thorwart2009enhanced,Prior2010,Chin:2013ia}.
These methods provide significant improvement in accuracy, including, for example, representation of non-Markovian effects,
but are generally considerably less efficient than the perturbative approaches.  For open loop control studies in which dynamical calculations need to be repeated
$\sim 10^3 - 10^4$ times per optimization target, it is essential to have an extremely efficient numerical simulation technique.  While recent developments in new approximate methods \cite{ritschel2011efficient} as well as fast implementations of exact methods using graphics processing units (GPUs) \cite{Kreisbeck2011} are significantly improving the ability to undertake efficient dynamical calculations and also to probe more realistic models of system-reservoir coupling, these methods are still typically two orders of magnitude slower than Redfield dynamics. The time constraints are further exacerbated when the goal is coherent control of a pump-probe or two-dimensional spectroscopic experiment, requiring dynamical calculation of higher order polarizabilities.

While direct optimization using more accurate dynamical descriptions could be feasible for future studies,
for this study we use Redfield theory in the secular approximation, which we believe is the best perturbative approach for modeling quantum dynamics in natural light harvesting systems.
We can evaluate Redfield dynamics for a single molecule of FMO in ${\sim 100}$ ms on a single modern CPU, compared to up to hours for more accurate methods \cite{Ish.Fle-2009b,ritschel2011efficient,huo2010iterative,thorwart2009enhanced,Prior2010,Chin:2013ia}).  Such rapid simulation of the dynamics is a necessary trade-off to ensure we can fully optimize candidate control pulses within a reasonable time frame.
It also allows exploration of the dependence of the coherent control upon the system Hamiltonian and the system-bath coupling.
Furthermore, one can subsequently verify the accuracy of the control schemes optimized in this fashion by carrying out spectroscopic simulations with one of the more advanced simulation methods using the ``Redfield optimized'' pulses, as we demonstrate in Sec.~\ref{sec:limits}.

The approach of secular Redfield theory is to treat the system-reservoir coupling $H_\text{S-R}$ to second order and to apply the Markov and secular approximations \cite{May2011}. This allows us to write an equation of motion for the electronic reduced density matrix $\rho$,
\begin{align}
	\frac{d\rho}{dt} &= -\frac{i}{\hbar} [H_\text{S} + H_\text{S-L}(t), \rho] + \mathcal{D} \rho
	\label{eq:EOM}
\end{align}
where the dissipation superoperator $\mathcal{D}$ can be written as a sum of components in the standard Lindblad form. We include the explicit time-dependence in $H_\text{S-L}$ to indicate that this is the only non-constant term (aside from $\rho$), which holds under the assumption that we can neglect modification of the dissipative dynamics due to strong field excitation \cite{Schirrmeister:1998gq}.
This description gives rise to qualitatively reasonable dynamics including quantum beats and relaxation to the proper thermal equilibrium, unlike other efficient simplified quantum master equations such as the Haken-Strobl
or pure dephasing model \cite{Haken1973,Cao2009}.
Employing the secular approximation guarantees completely positive dynamics, at the price of neglecting environment induced dynamical coherence transfer \cite{Panitchayangkoon2011}.

To efficiently simulate light-matter interactions, we use the rotating wave approximation, which allows us to ignore the contribution of very quickly oscillating and non-energy conserving terms \cite{Shapiro2012}.
To do so, we switch to a rotating frame with frequency $\omega_0$, typically chosen to match the carrier frequency of the applied field. The dynamics are given by replacing each Hamiltonian $H$ with its rotating equivalent $\widetilde{H}$. The only terms in Eqs.~\eqref{eq:H_full}-\eqref{eq:H_S-L} which are not equivalent to those in the non-rotating frame is the system Hamiltonian $\widetilde{H}_S$, which in the rotating frame has the transition energies $\widetilde{\mathcal{E}}_n = \mathcal{E}_n - \hbar \omega_0$, and the system-light interaction
\begin{align}
	\widetilde{H}_\text{S-L} = \sum_n (\vec{d}_n \cdot \hat{e})~ a_n \widetilde{E}(t) + h.c.
\end{align}
The electric field in the rotating frame, $\widetilde{E}(t)$, is related to the full electric field by $\vec E(t) = \hat e \widetilde{E}(t) e^{i\omega_0 t} + c.c.$, where $\hat e$ is a unit vector in the direction of the polarization of the field. Since the Redfield dissipation superoperator $\mathcal{D}$
is time-independent, to determine the electronic density matrix $\rho$ resulting from a control field we may numerically integrate Eq.~\eqref{eq:EOM} upon substituting $\widetilde{H}_\text{S}$ and $\widetilde{H}_\text{S-L}$ for $H_\text{S}$ and $H_\text{S-L}$, respectively. Weak fields also allow us to perform efficient averages over all molecular orientations \cite{Craig1984}.

\section{Optimal control methods}
\label{sec:methods}

We focus here on optimizing state preparation under weak field excitation, the experimentally relevant regime for ultrafast spectroscopy.
Spectroscopic measurements usually cannot determine the fraction of an ensemble sample that is excited, so we employ the normalized excited state density matrix given by
\begin{align}
	\rho^\prime = \frac{\rho_e}{\tr \rho_e}, \label{eq:rho_prime}
\end{align}
where $\rho_e$ denotes the projection of the density matrix onto the single excitation subspace. In the weak field regime, $\rho^\prime$ will not be sensitive to the overall pump amplitude, since the total excited state population will remain small.

Our control objectives can all be expressed in terms of this normalized excited state density matrix $\rho^\prime$.
Our first state preparation goal is localization of the excitation on a pigment $n$:
\begin{align}
\max C^{\text{site}}_n(t) \equiv \max \bra{n}\rho^\prime (t) \ket{n}
\label{eq:c_site}
\end{align}
for $n \in \{1,2,\ldots,7\}$, with $a_n\dg a_n \ket{m} = \delta_{mn} \ket{m}$. The second goal we consider is the preparation of an excitonic state:
\begin{align}
\max C^{\text{exc}}_\alpha(t) \equiv \max \bra{\alpha}\rho^\prime (t) \ket{\alpha}
\label{eq:c_exc}
\end{align}
for $\alpha \in \{1,2,\ldots,7\}$, where the excitonic states $\ket{\alpha}$ are the eigenstates of $H_\text{S}$, and we choose to order them by energy so that exciton $\alpha=1$ is the lowest energy one. The third goal is the maximization of an excitonic coherence:
\begin{align}
\max C^{\text{coh}}_{\alpha,\beta}(t) \equiv \max |\bra{\alpha}\rho^\prime (t) \ket{\beta}|
\label{eq:c_coh}
\end{align}
for $\alpha,\beta \in \{1,2,\ldots,7\}$. The time $t$ in all these cost functions is restricted to times following the end of the shaped control pulse. In addition to these state population targets, in section \ref{sec:authentication} we describe and carry out optimization of another cost function of $\rho^\prime(t)$ that is directly related to experimentally measurable spectroscopic signatures.

\begin{figure}[t]
	\includegraphics{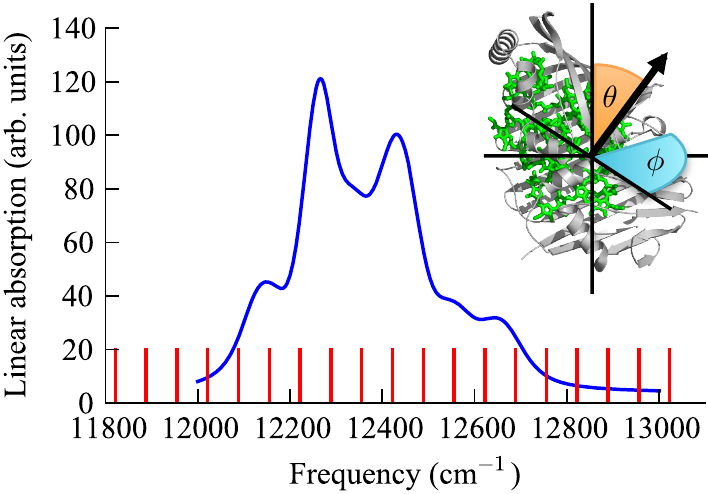}
	\caption{\label{fig:crab} Linear absorption spectrum of FMO together with representation of the values used in the first pulse parameterization for controlling excitation of FMO, Eq.~\eqref{eq:CRAB}.  The blue solid line shows the FMO linear absorption spectrum. The vertical red lines are located at the chosen frequencies $\omega_k$ of Eq.~\eqref{eq:CRAB}.  The schematic inserted at top right shows the angles defining the laser polarization relative to the fixed molecular frame. }
\end{figure}

We employ two types of pulse parameterizations in this work, optimizing the parameterized control fields in each case with standard numerical optimization techniques (see below).
To first determine the theoretical limits of control without taking any constraints on physical realization of the control fields, we use an adaptation of the chopped random basis (CRAB) scheme \cite{caneva2011chopped}, which was recently used to optimize state preparation for FMO \cite{Caruso:2012hv}.
This control field is defined in the time domain and is of the form
\begin{align}
	E(t) = f_\text{cutoff}(t) \sum_k^N \left(A_k + i B_k \right) e^{i \omega_k t},
	\label{eq:CRAB}
\end{align}
for $N$ fixed frequencies $\omega_k$ and $2N$ real valued optimization parameters $A_k$ and $B_k$.
This parameterization allows variation of both amplitude and phase of the control field components.
In our case, we choose $N=19$ frequencies $\omega_k$ at equal intervals spanning the FMO absorption spectrum.
The cutoff function $f_\text{cutoff}(t)$ is chosen as a step function that restricts the control fields to a particular total pulse duration $T$.
For complexes with fixed orientation (in contrast to optimization over an ensemble of orientations), we also include parameters $\phi$ and $\theta$ to choose the optimal polarization $\hat e$ of the control field. These optimization parameters are illustrated in Figure \ref{fig:crab}.

The parametrization given by Eq.~\eqref{eq:CRAB} creates pulses with different overall energies and energy distributions, but ``coherent control'' is in some cases understood as referring to exploiting the coherent nature of the laser light itself \cite{Shapiro2012}. For this reason, some initial experimental demonstrations of coherent control have stressed the sensitivity of control to the \emph{phase} of the control field \cite{Prokhorenko2006}. Thus, we also consider a second pulse parametrization suited for coherent control within an experimentally constrained phase-only control scenario.
Here we define the control field in the frequency domain and each frequency is assigned a variable phase term which is parameterized as a polynomial function of the frequency:
\begin{align}
	E(\omega) = A(\omega) \exp\left[i \sum_{k=2}^{N+1} C_k (\omega - \omega_0)^k \right].
	\label{eq:poly-phase}
\end{align}
The quantity $A(\omega)$ denotes the unshaped amplitude profile of the control field, which is fixed by the laser setup, $\omega_0$ is the central frequency of the laser and $C_k$ is a real valued optimization parameter. We neglect terms with $k < 2$, in order to remove global phase shifts ($k=0$) as well as time-translations ($k=1$). The surviving terms $k \geq 2$ correspond to linear ($k=2$) and higher-order ($k>2$) chirps in the frequency domain.
In this work we have chosen the amplitude $A(\omega)$ to be represented by a Gaussian function, with amplitude fixed to \SI{\sim e7}{\volt\per\meter} in the time-domain and standard deviation \SI{225}{cm^{-1}}, corresponding to a full-width-at-half-maximum of \SI{55}{fs} in the time domain.
In this paper, we restrict ourselves to $N=10$ terms.
To simulate hypothetical pulses matching those produced by an ultrafast pulse shaper \cite{SchlauCohen:2012du,Turner:2011hr}, we apply this field in the frequency domain to 600 points stretching between $\pm 3$ standard deviations, and obtain the time-domain pulse in the rotating frame from the inverse Fourier transform. Cutting off frequencies more than 3 standard deviations away from the central frequency does result in some small artifacts visible in the time-domain. We then use the following two-step heuristic to determine the final time for the pulse: we set to zero the pulse field at all times with amplitude less than 0.005 times the maximum amplitude and then set the final time as the point at which 99\% of the overall integrated total pulse amplitude has passed.
For optimization of oriented complexes, the relative polarization $\hat e$ is specified by angles $\theta$ and $\phi$ as above.

To optimize our pulses over the parameters in Eqs.~\eqref{eq:CRAB} and \eqref{eq:poly-phase} we use two optimization routines, the Subplex direct search algorithm \cite{Rowan1990} and the Covariance Matrix Adaptation Evolution Strategy (CMA-ES)  \cite{Hansen:2006fy} genetic algorithm. Subplex is a local, derivative free search algorithm that has been previously used with control fields using the CRAB parametrization \cite{caneva2011chopped,Caruso:2012hv}. CMA-ES is a genetic algorithm designed for solving difficult continuous optimization problems while employing only a few tunable hyper-parameters (here we set the population size to 50 and the initial standard deviation to 1).
We limited each optimization algorithm to a maximum of \num{10000} function evaluations with a single initial condition, $A_k=B_k=1$ for Eq.~\eqref{eq:CRAB} and $C_k=0$ for Eq.~\eqref{eq:poly-phase}. Interestingly, we found that the best optimization algorithm depended on the particular optimization target (e.g., Subplex worked better than CMA-ES for 69\% of the cases depicted in Fig.~\ref{fig:venn-diagram}), so we used the best result from optimizations with both algorithms. With more function evaluations available, more random initial conditions could be used with Subplex (as in Ref.~\cite{Caruso:2012hv}) and population sizes could increased for CMA-ES.
Because both algorithms struggle to optimize over very rough control landscapes, when implementing using the pulse parameterization in Eq.~\eqref{eq:CRAB}, we choose the total pulse duration with a separate optimization, by performing a grid search over all pulse durations between \SI{100}{fs} and \SI{500}{fs} at increments of \SI{50}{fs}. In practice, we find this significantly increases the consistency and quality of our optimization results compared to including the pulse duration as a search parameter.
In addition, to guarantee that our optimal pulses are still in the weak field regime with Eq.~\eqref{eq:CRAB}, we subtract from each objective function a smooth cutoff function of the form $\Theta(x - \alpha) (x - \alpha)^2$, where $x$ is the total excited state population and $\alpha=0.01$ is set as the maximum acceptable excited state population. Note that for an unshaped Gaussian pulse (i.e., Eq.~\eqref{eq:poly-phase} with all $C_k=0$), this maximum excited state population of 1\% corresponds to a maximum electric field amplitude in the time domain of no more than \SI{1.2e7}{\volt\per\meter}.

To match experimental conditions, in addition to optimizing over pulse parameters, in most of our optimizations we average over ensembles of complexes characterized by orientational and/or energetic (static) disorder. For these calculations, we use two approaches. To average over the orientations, we use the fact that the isotropic average of the excited state density matrix under weak fields is equal to the average of the x, y and z polarization orientations (see Appendix \ref{sec:orientational-average}). Accordingly, we can obtain the exact orientational average at the cost of only a factor of 3 times more function evaluations. Unfortunately, there is no such shortcut for averages over energetic disorder. In such cases, we perform optimizations on the average of 10 randomly (but consistently) chosen samples, and still limit ourselves to the fixed computational cost of \num{10000} function evaluations by now allowing for no more than \num{1000} function evaluations. To determine final optimal results in the presence of static energetic disorder, we average over ensembles of \num{1000} samples. For these simulations, we average the excited state density matrix $\rho_e$ in the site basis before inserting it into the objective functions given by Eqs.~(\ref{eq:rho_prime}--\ref{eq:c_coh}).

\section{Limits of coherent control for initial state preparation in FMO}
\label{sec:limits}

An interesting question to consider before investigating control pulses designed for specific target states, is whether or not the FMO system can be prepared in any arbitrary state, i.e., whether FMO is controllable.  Formally, a closed quantum system is completely controllable if control fields can be used to generate any arbitrary unitary operation \cite{Schirmer2001}. Complete controllability implies that it is possible to prepare arbitrary pure states starting from any initial pure state, which in our case is the ground state. The controllability of a closed quantum system can be analyzed algebraically, by examining the Lie algebra generated by the system and control Hamiltonians.
Accordingly, we consider the following highly idealized scenario: the electronic degree of freedom for a single FMO molecule without coupling to vibrations and restricted to at most one electronic excitation. Inter-pigment and pigment-protein interactions break the degeneracy of the excited states, so for the system without dissipation there actually is a constructive algorithm for determining arbitrary unitary controls \cite{Ramakrishna:2000td}. We confirmed this by running the algorithm of Ref.~\cite{Schirmer2001} to verify that we do indeed have complete controllability for formation of exciton states in the FMO complex under the combination of the electronic system Hamiltonian $H_S$ and the light-matter interaction $H_{S-L}$ with any arbitrarily chosen single polarization of light.

In contrast, there are few rigorous results known for determining the controllability of open quantum systems \cite{Dirr:2009}.
Accordingly, in our calculations below for realistic experimental conditions (weak fields, bulk samples with orientational and static disorder, finite temperatures) we assess the feasibility landscape of both arbitrary (amplitude/phase) and phase-only control in a brute-force manner, namely by evaluating the effectiveness of candidate control pulses for various targets. Our constraints are chosen to reflect the current experimental situation and include the use of weak fields, averaging over all orientations, averaging over static disorder and decoherence at \SI{77}{K}. Averaging over orientations and over static disorder would not be necessary in single complex spectroscopy experiments, which may become feasible in the foreseeable future (see below).
We have carried out optimizations to both maximize and to minimize population on specific sites and excitonic states.

\begin{figure}[tb]
	\includegraphics[width=3.375in]{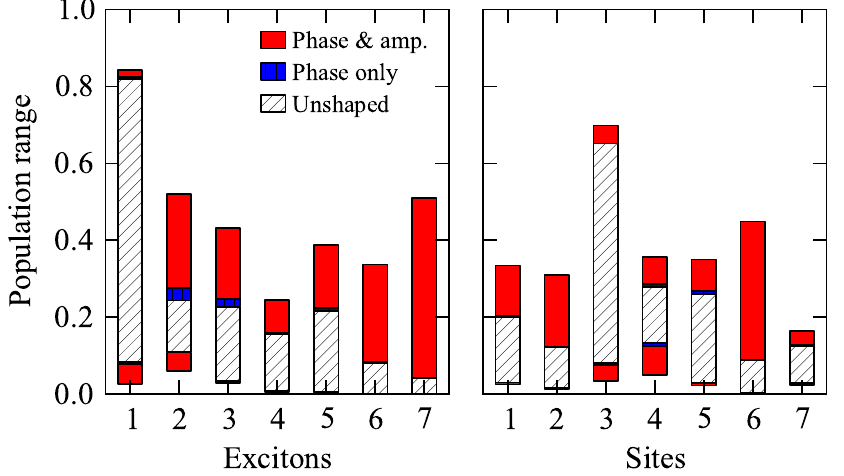}
	\caption{\label{fig:pop-control}Population control results for all targets (site and exciton populations).
	Blue and red bars indicate the minimum-to-maximum populations achievable with phase only control and with phase plus amplitude control, respectively.  The white bars indicate the minimum-to-maximum achievable by using unshaped pulses together with a variable time delay following the pulse (see text). Phase only control is a special case of phase and amplitude control, and unshaped pulses are a special case of shaped pulses, so the red area is actually inclusive of the entire range shown and the blue area is inclusive of the white bar. Exciton 1 has the lowest energy and exciton 7 the highest energy.}
\end{figure}

The results of these optimizations with amplitude and phase control [Eq.~\eqref{eq:CRAB}] and with phase-only control [Eq.~\eqref{eq:poly-phase}], are summarized in Figure \ref{fig:pop-control} and compared with the corresponding populations achieved using unshaped pulses.
The latter is the set of populations realized by a fixed ``unshaped'' Gaussian pulse with a variable time-delay after the end of the pulse.
The variable time-delay simply allows incorporation of the inherent relaxation in the system, which can aid in optimizing some goals (e.g., preparation of the lowest energy exciton, or of site 3, on which the lowest exciton is mostly localized).
These results show that under experimental conditions corresponding to bulk samples of FMO complexes in solution at liquid nitrogen temperatures, we are in a coherent control situation where fidelity values are more similar to those achieved in typical reactive chemical control situations than the values required for quantum information processing, underscoring once again the critical role of the open quantum system dynamics.

\begin{figure}[]
	\includegraphics[]{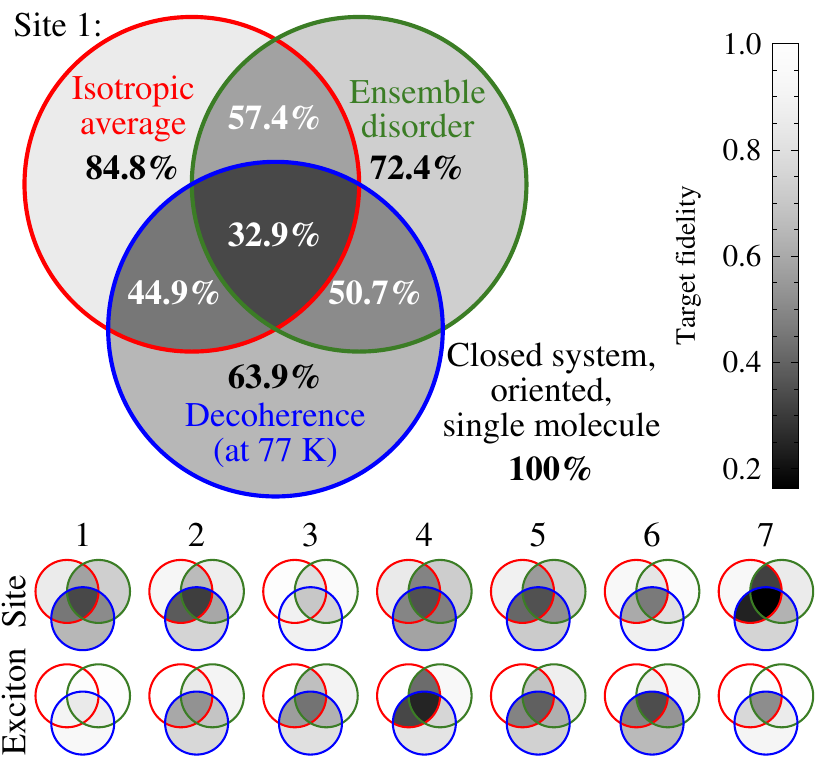}
	\caption{\label{fig:venn-diagram}Venn diagrams indicating control fidelity for optimization under different experimental constraints with joint phase and amplitude control.
	The outer region refers to the closed quantum system constituted by the excitonic Hamiltonian without coupling to the protein environment.  Here the excitonic system is completely controllable and the fidelity of preparing any state is 100\%.  The circles enclosed by colored lines indicate the fidelity achieved when we add averaging over orientation of the FMO complex (red circles), averaging over site energy disorder (green circles) and adding the coupling to the protein environment at a finite temperature (blue circles).
	(Above) Venn diagram for maximum population on site 1. (Below) Venn diagrams for all site and exciton targets.}
\end{figure}

We now consider the effect of each of three major constraints on the achievable fidelity of control, both singly and in all possible combinations.
We identify these three primary constraints as the following: the isotropic average over all molecular orientations, the ensemble average over static disorder of the transition energies $\mathcal{E}_n$ and the effects of decoherence due to the interaction with the environment (represented by a bath of phonon modes) at \SI{77}{K}. The independent and cumulative effect of these constraints are illustrated by the Venn diagrams in Figure \ref{fig:venn-diagram} and the values are given in Table \ref{tab:venn-diagram}.
\begin{table}[]
	\begin{ruledtabular}
	\begin{tabular}{cr|cccccccc}
	& & \multicolumn{4}{c|}{Unitary} & \multicolumn{4}{c}{Decoherence (at 77 K)} \\ \cline{3-10}
	& & \multicolumn{2}{c|}{Single} & \multicolumn{2}{c|}{Ensemble} & \multicolumn{2}{c|}{Single} & \multicolumn{2}{c}{Ensemble} \\ \cline{3-10}
	\multicolumn{2}{c|}{Target} & \multicolumn{1}{c|}{Orient.} & \multicolumn{1}{c|}{Iso.} & \multicolumn{1}{c|}{Orient.} & \multicolumn{1}{c|}{Iso.} & \multicolumn{1}{c|}{Orient.} & \multicolumn{1}{c|}{Iso.} & \multicolumn{1}{c|}{Orient.} & \multicolumn{1}{c|}{Iso.} \\ \hline
	\multirow{8}{*}{\begin{sideways}\parbox{1.0cm}{\centering Site}\end{sideways}}
	& 1 & 100.0 & 84.8 & 72.4 & 57.4 & 63.9 & 44.9 & 50.7 & 32.9 \\
	& 2 & 100.0 & 91.1 & 86.4 & 65.2 & 74.3 & 37.6 & 58.7 & 30.6 \\
	& 3 & 100.0 & 97.2 & 94.0 & 79.1 & 87.8 & 87.9 & 74.9 & 69.4 \\
	& 4 & 100.0 & 83.9 & 71.8 & 48.2 & 57.5 & 48.7 & 47.3 & 35.2 \\
	& 5 & 100.0 & 88.0 & 74.5 & 55.0 & 71.2 & 40.7 & 51.7 & 35.2 \\
	& 6 & 100.0 & 92.4 & 94.9 & 81.6 & 87.3 & 59.6 & 80.1 & 45.7 \\
	& 7 & 100.0 & 92.6 & 84.7 & 31.3 & 74.5 & 23.8 & 62.2 & 16.3 \\ \cline{2-10}
	& Mean & 100.0 & 90.0 & 82.7 & 59.7 & 73.8 & 49.0 & 60.8 & 37.9 \\ \hline
	\multirow{8}{*}{\begin{sideways}\parbox{1.0cm}{\centering Exciton}\end{sideways}}
	& 1 & 100.0 & 100.0 & 96.6 & 95.3 & 92.4 & 89.9 & 86.5 & 84.2 \\
	& 2 & 100.0 & 100.0 & 89.4 & 83.6 & 76.1 & 59.0 & 68.6 & 52.5 \\
	& 3 & 100.0 & 100.0 & 89.3 & 66.7 & 79.4 & 55.8 & 62.4 & 44.0 \\
	& 4 & 100.0 & 99.9 & 92.2 & 41.9 & 81.5 & 32.7 & 75.3 & 24.4 \\
	& 5 & 100.0 & 99.9 & 86.9 & 68.1 & 73.1 & 48.3 & 61.5 & 39.0 \\
	& 6 & 100.0 & 100.0 & 92.7 & 87.2 & 64.9 & 49.1 & 53.5 & 34.1 \\
	& 7 & 100.0 & 100.0 & 98.4 & 97.2 & 90.5 & 76.0 & 84.9 & 51.1 \\ \cline{2-10}
	& Mean & 100.0 & 100.0 & 92.2 & 77.1 & 79.7 & 58.7 & 70.4 & 47.1 \\
	\end{tabular}
	\end{ruledtabular}
	%BW improved caption for Referee 2
	\caption{\label{tab:venn-diagram} Optimized fidelities for the calculations that are illustrated schematically in Fig.~\ref{fig:venn-diagram}. (Top panel) Results for optimization of site populations.  (Bottom panel) Results for optimization of exciton populations.  The electronic Hamiltonian $H_\text{S}$ is completely controllable under action of the light-matter interaction $H_\text{S-L}$ with a single polarization: hence all values in the first column of the table (unitary, single complex, fixed orientation with respect to the laser polarization) are equal to 100\%.  Subsequent columns to the right of this show the effects of orientational averaging, ensemble (energy disorder) averaging, and the addition of decoherence (dephasing and dissipation) at 77 K.}
\end{table}
We have restricted the depictions here to just the results for maximizing each exciton or site population and for joint phase and amplitude control using Eq.~\eqref{eq:CRAB}.

Although it is clear from the lower panels of Figure~\ref{fig:venn-diagram} that the precise results of optimization are different for each target state, we can nevertheless extract a number of striking trends. For example, for most targets, the single largest limiting factor to controllability is decoherence.
Also, in general, it is easier to achieve single exciton rather than site localized target states, as indicated by the average control fidelities over all sites or all excitons in Table \ref{tab:venn-diagram}.  Site 3 is an exception to both these trends, which is readily rationalized by recognizing that the lowest energy exciton (exciton 1) is primarily localized on site 3.  Indeed, it is notable that optimization both of population on site 3 and in exciton 1 are relatively robust to all three of the constraints. We can also see that averaging over orientations without decoherence is evidently a more severe constraint for localized site target states than for single exciton target states. This can be rationalized as reflecting a simple physical strategy for targeting excitonic states, namely, to drive the system for a long time at the proper transition energy. In most cases, it is clear that the difficulty of control under realistic conditions stems from the combination of two or more of these limiting factors, which together rule out many intuitive control strategies.

\begin{figure*}[]
	\includegraphics[]{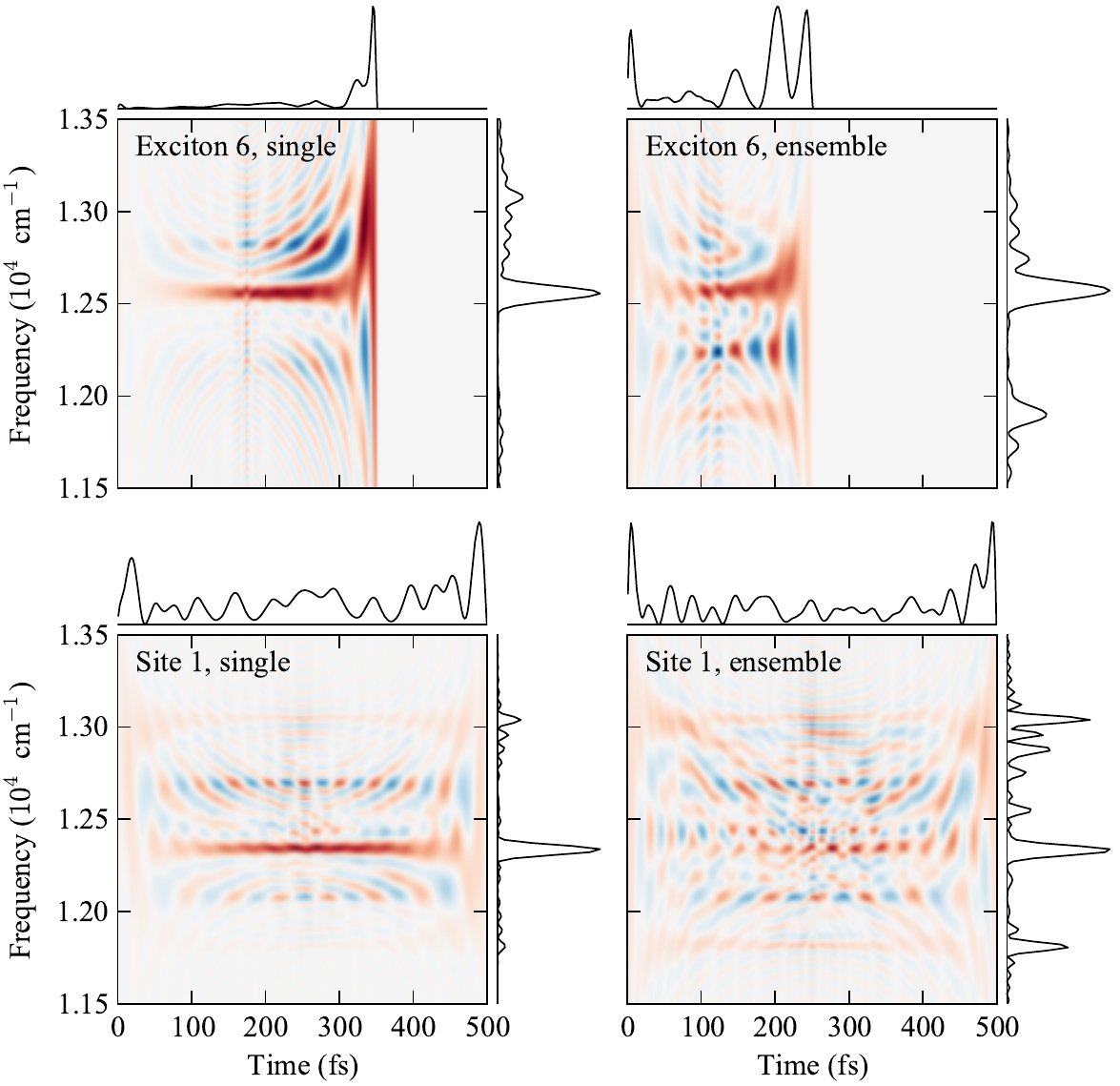}
	\caption{\label{fig:pulses}Optimal pulses to maximize the population of the isotropic average on exciton 6 (top) or site 1 (bottom) in the presence of decoherence, both without (left) and with (bottom) static disorder. The central plot of each pulse is of the Wigner spectrogram.
	Red (blue) indicates positive (negative) values. Top and right plots show the marginal time and frequency distributions $|E(t)|^2$ and $|E(\omega)|^2$.}
\end{figure*}

The optimal pulse shapes for selected instances of site and exciton target states, with and without ensemble averaging over static disorder, are shown in Figure \ref{fig:pulses}.
We illustrate these pulses by plotting their Wigner spectrograms \cite{Mukamel:1996ts}, given by the expression
\begin{align}
W(t, \omega) &= \int_{-\infty}^\infty d\tau E^\ast(t - \tau/2) E(t + \tau/2) \exp(i\omega\tau),
\end{align}
where $E(t)$ is the pump field and $\ast$ denotes the complex-conjugate.
The Wigner spectrogram can be interpreted as a type of joint time-frequency distribution, with the desirable properties,
\begin{align}
|E(t)|^2 &= \int \frac{d\omega}{2\pi} W(t, \omega) \\
|E(\omega)|^2 &= \int dt W(t, \omega).
\end{align}
Recall that because we are in the the weak-field regime and normalize all objective functions by the excited state population, our results are independent of the overall amplitude of the pump pulse, so there is no need to provide an absolute scale for the control field amplitude.
As Figure \ref{fig:pulses} shows, optimal pulses for
targeting states in the presence of static disorder generally require a more complicated frequency distribution and a broader band of excitation frequencies. This is particularly evident from the optimal pulses for creating an excitation in a single excitonic state, where the optimal pulses are generally peaked at the transition energy of the desired state.
The optimal pulses for preparing an excitation in exciton 6 (top panels) illustrates also the complementary trend that in the presence of static disorder the optimal pulses are usually shorter in time. Not surprisingly, pulses designed with optimal polarizations (i.e., without the average over all orientations), are even shorter since these optimizations can make use of polarization in addition to or instead of excitation frequency to selectively target specific states.

\begin{figure}[]
	\includegraphics{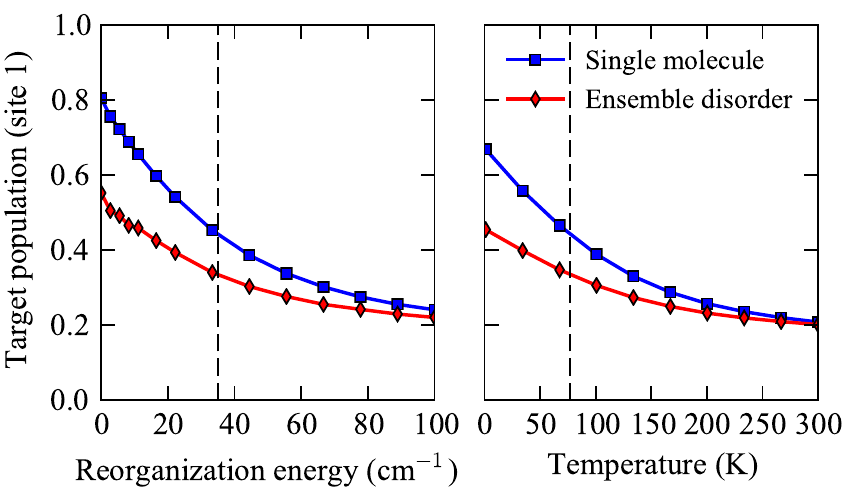}
	\caption{\label{fig:modify-decoh}Maximum target population (site 1) as a function of the bath reorganization energy (left) and temperature (right), hence modifying the amount of decoherence in the excitation dynamics. The results are shown in the orientational average with and without including ensemble disorder. The standard bath parameter choices, used when not modifying the decoherence rate, are indicated with the vertical dashed lines. Left: Temperature is fixed at \SI{77}{K}. Right: Reorganization energy is fixed at \SI{35}{\per\centi\meter}.
	}
\end{figure}

To explore the influence of decoherence in more detail, we consider the independent effects of adjusting the temperature of the bath and the reorganization energy. Figure \ref{fig:modify-decoh} shows the error for maximizing population in one specific site, as a function of the reorganization energy (left panel) and as a function of the bath temperature (right panel).
We see that although lowering the bath temperature to near zero does make it easier to target site 1, this is not as powerful a control knob as removing the system-bath coupling (i.e., setting the bath reorganization energy to zero).

\begin{figure}[]
	\includegraphics{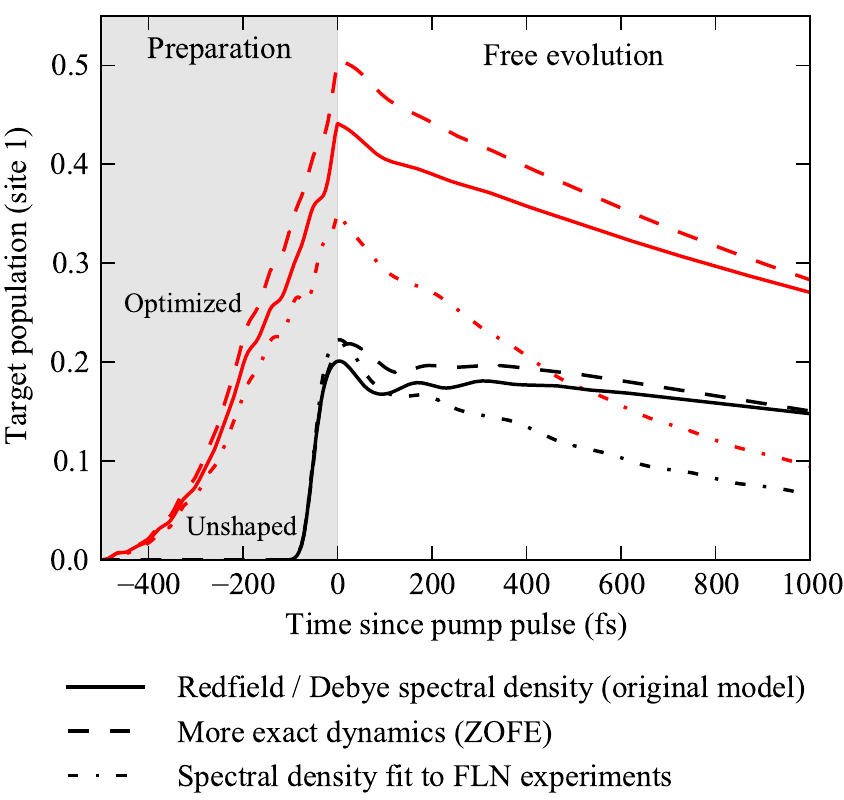}
	\caption{\label{fig:alternative-dynamics}Target population for site 1 shown as a function of time for dynamical simulations with three different dynamical methods using the ``Redfield optimized'' pulses in all cases. The dynamical models are described in the text.
	Each of these simulations makes a orientational average but does not include ensemble disorder.
	}
\end{figure}

We now consider the robustness of our optimized control pulses to the necessary limitations in the dynamical model for the coherent control calculations that was discussed earlier (Sec.~\ref{sec:introduction}).
Figure \ref{fig:alternative-dynamics} compares the population dynamics at a target site for optimal and unshaped pulses employing the Redfield secular dynamical model used for all $\sim \num{10000}$ calculations during the optimization (solid black and red lines), with the population dynamics approximated by the non-Markovian quantum state diffusion method under the zeroth order functional expansion (ZOFE) \cite{ritschel2011efficient} (dashed black and red lines). The ZOFE approximation is a relatively computationally efficient technique for modeling dynamics in the FMO complex that has also been shown to have excellent agreement with the results of theoretically exact models, such as the 2nd-order time-nonlocal cummulant expansion, at a fraction of the computational expense. Here we use the formulation and time-correlation function from Ref.~\citenum{ritschel2011efficient}, which was fit specifically to the same Debye spectral density at \SI{77}{K} that we use for our Redfield model.   However, our calculations with this method are still $\sim 100 \times$ slower than for Redfield theory, which makes performing optimizations under ZOFE significantly more expensive.
The results show that the population enhancement due to ``Redfield optimized'' control pulses is robust to the different levels of accuracy in the dynamical simulation, performing essentially equally well with the Redfield and ZOFE dynamics. In fact, in this particular example the target population is even larger under ZOFE dynamics, even though the optimization was done for our original Redfield model.

Given this demonstrated robustness of the optimization with respect to the underlying dynamical model, we can now investigate robustness with respect to the form of the spectral density. In Figure \ref{fig:alternative-dynamics} we compare population dynamics under the Debye spectral density (solid black and red lines) to dynamics under a spectral density derived from Fluorescence Line Narrowing (FLN) experiments \cite{Adolphs2006}, for which we model dynamics with secular Redfield dynamics neglecting the imaginary part of the time-correlation function (dot-dashed black and red lines).
The later spectral density has been used in a number of recent studies of dynamics in the FMO complex \cite{Nalbach:2011ef, Kreisbeck:2012bl, Chin:2013ia}.
We see that the optimization is again relatively robust to these differences in system-bath coupling although, not surprisingly, the free evolution after the pulse is switched off does depend on the form of the spectral density, with that of Ref.~\cite{Adolphs2006} showing systematically faster decay of the population at site 1.
This is due to the higher values of this spectral density than the Debye spectral density at most transition energies between exciton states \cite{Nalbach:2011ef}, which, within Redfield theory, is directly proportional to the energy transfer rate.

\section{Authentication of coherent control in pump-probe spectra}
\label{sec:authentication}

To experimentally verify the excited state that is created by a control pulse, we propose to use ultrafast pump-probe spectroscopy. In pump-probe spectroscopy, a pump pulse excites the system, followed by a probe pulse at a controlled delay time.
Although there are a number of ultrafast spectroscopies that may be used for probing light harvesting systems \cite{Mukamel1995}, pump-probe spectroscopy is particularly well suited for verifying state preparation since its signal can be formally interpreted as a function of the full excited state created by the pump pulse \cite{Hoyer:2012ud}.
In the limit of an impulsive probe pulse, provided that we are in the weak field regime and that there is no time overlap between the pump and probe pulses, the frequency resolved pump-probe signal is given by
\begin{align}
	S(\omega, T) \propto \text{Re}\left\{ \tr [ \mu G(\omega) [\mu^\dagger, \rho^{(2)}_\text{pu}(T)]] \right\},
	\label{eq:pump-probe-general}
\end{align}
where $\omega$ is the probe frequency and $T$ is the time delay of the probe relative the end of the pump probe  \cite{Hoyer:2012ud}.
We emphasize that one cannot verify state preparation with this approach before the end of the pump pulse. The operator $G(\omega)$ is the Fourier transform of the retarded material Green function,
$\mu$ is the annihilation component of the dipole operator $\vec\mu$ in the direction of the probe field and $\rho^{(2)}_\text{pu}(T)$ is the second order contribution to the density matrix in terms of the pump pulse, restricted to the 0-1 excitation manifold. For a Markovian bath and weak excitation, this density matrix $\rho^{(2)}_\text{pu}(T)$ can be written as a linear function of the excited state density matrix $\rho_e(T)$ \cite{Hoyer:2012ud}. Accordingly, we can write Eq.~\eqref{eq:pump-probe-general} in the form
\begin{align}
	 S(\omega, T) \propto \text{Re} \left\{ \tr [ \mathcal{P}(\omega) \rho_e(T)] \right\}
	 \label{eq:signal}
\end{align}
where $\mathcal{P}(\omega)$ is a (linear) superoperator in Liouville space.
This equation still holds even for a randomly oriented ensemble for a pump-probe experiment performed in the magic-angle configuration \cite{Hochstrasser2001}, provided that $\mathcal{P}(\omega)$ and $\rho_e$ are replaced by their isotropic averages (see Appendix \ref{sec:orientational-average}). We then efficiently perform each of these independent isotropic averages by averaging over the x, y and z pump (or probe) polarizations \cite{Craig1984}.

The ideal authentication procedure for any quantum control experiment is  quantum state tomography, in which each element in the density matrix is determined by repeated measurements on an ensemble of identically prepared systems \cite{Nielsen2000}. It is possible to construct protocols for state tomography of excitonic systems that are based on pump-probe spectroscopy \cite{Yuen-Zhou2011, Hoyer:2012ud}, but the uncertainty concerning Hamiltonian parameters, together with the averaging over energetic disorder and orientations associated with ensemble measurements, make this an unrealistic target for systems with as many pigments as FMO \cite{Hoyer:2012ud} Accordingly, here we consider two options for authenticating control that may be realized with current experimental technology.
The first is to extend the duration of coherent electronic beating and the second is to generate a novel signature in the pump-probe signal.
We note that with a frequency resolved signal as given by Eq~\eqref{eq:pump-probe-general} there are no gains from optimizing the probe pulse, as long as it is faster than the dynamics of interest and has non-zero amplitude at the frequencies of interest.

\begin{figure}[]
	\includegraphics[]{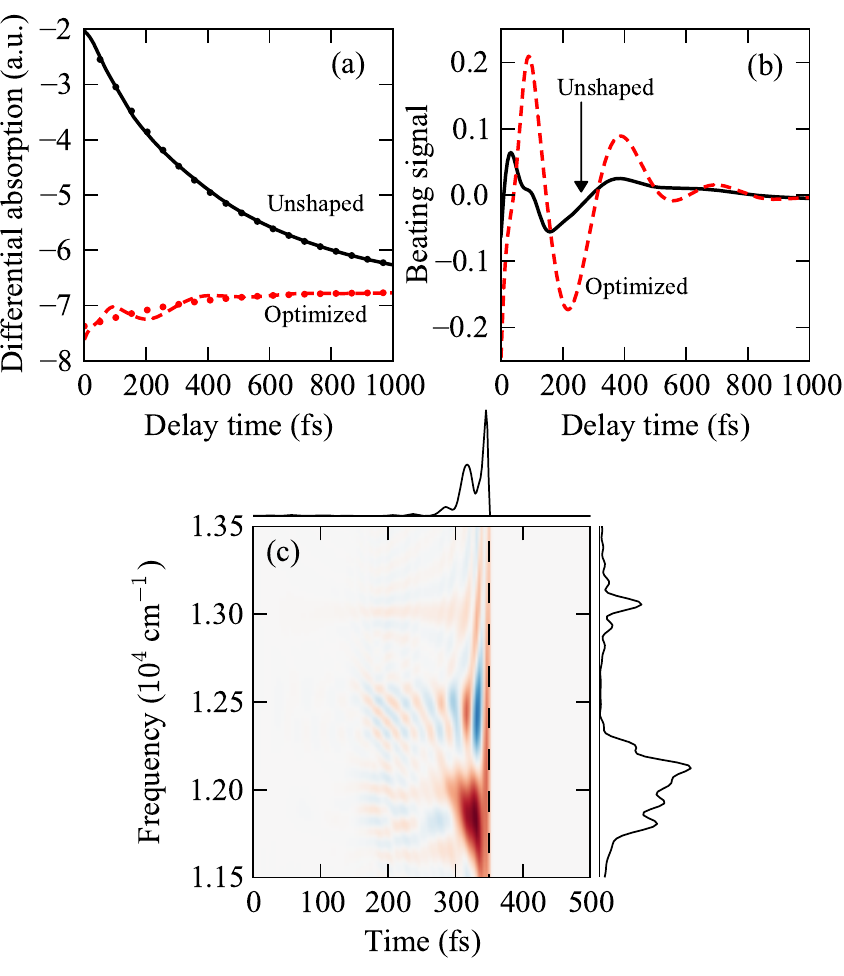}
	\caption{\label{fig:coherent-beating}Enhanced coherent beating in the pump-probe signal. (a) Differential absorption [the pump-probe signal given by Eq.~\eqref{eq:pump-probe-general}] at a probe frequency of \SI{12200}{cm^{-1}} following unshaped
	(solid blue line) or optimized pump (dashed red line) pulses, normalized by the total excited state population following the pump pulse. The probe frequency was chosen as an arbitrary value, in the range of the exciton 1 and 2 transition energies where the oscillatory signal from the optimized pump is clear. Dots show a double exponential fit for each signal of the form $A e^{-a t} + B e^{-b t} + C$, where $A$, $B$, $C$, $a$ and $b$ are constants chosen to minimize the least squares distance between the fit and the differential absorption signal. (b) Beating portion of the signal shown in panel (a) after subtracting the double exponential fits, $A=\SI{3.13}{cm^{-1}}, B=\SI{1.79}{cm^{-1}}, C=\SI{-6.89}{cm^{-1}}, a=\SI{3.11}{ps^{-1}}, b=\SI{1.29}{ps^{-1}}$ for the unshaped pulse and $A=\SI{44.8}{cm^{-1}}, B=\SI{-45.4}{cm^{-1}}, C=\SI{-6.76}{cm^{-1}}, a=\SI{5.57}{ps^{-1}}, b=\SI{5.52}{ps^{-1}}$ for the optimized pulse.
	(c) Wigner spectrogram showing the time-frequency distribution of the optimized control pulse, depicted in the same style as the pulses in Fig.~\ref{fig:pulses}.}
\end{figure}

Although the evidence for electronic coherence today is more clearly evident in 2D photon-echo spectroscopy \cite{Eng.Cal.etal-2007, Pan.Hay.etal-2010}, the first evidence for quantum beats in excitonic dynamics of the FMO complex were actually found with pump-probe spectroscopy \cite{Savikhin1997}.
Under secular Redfield dynamics, the excitonic coherence terms are decoupled from all other dynamics and decay exponentially.
Hence to maximize the duration of quantum beating within this description of the dynamics, we merely need to maximize the absolute value of the coherence terms in the excitonic basis.  However, predicting the visibility of electronic quantum beats is complicated by the relative orientations of the relevant transition dipole moments, as well as by the need to observe any oscillations before they decay. Accordingly, we demonstrate here the result of numerically maximizing the excitonic coherence $|\rho_{12}|$ between the lowest two exciton energies, since observations of quantum beats in pump-probe spectra have previously been ascribed to this coherence \cite{Savikhin1997}. (We do not optimize for the ensemble value of this coherence, because the ensemble average of excitonic coherences in the eigenbasis of the Hamiltonian without static disorder do not necessarily decay to zero at long delay times.)
For this optimization we use the unrestricted parametrization of Eq.~\eqref{eq:CRAB}, in order to determine the limits of maximizing this signal. We note that the aforementioned experimental pump-probe study did use a form of \emph{incoherent} control to maximize the this signal, by scanning the central frequency of the pump pulse \cite{Savikhin1997}. Our results show that we can increase the magnitude of the excitonic coherence $|\rho_{12}|$ by a factor of up to 8.6x compared to an unshaped Gaussian pump.
Figure~\ref{fig:coherent-beating} illustrates the optimal pulse and its signature on the pump-probe spectra including ensemble disorder.
As shown in panel (a), the increased coherence results in increased oscillatory features that will be visible in a magic angle pump-probe experiment, particularly after subtracting away the non-oscillating contribution to the signal [panel (b)]. This subtraction of the non-oscillating signal is similar to the technique used
in \cite{Pan.Hay.etal-2010} to estimate the timescale of quantum beats in a 2D photon-echo experiment. Even after including static disorder, the beating signal following the optimized pump shows oscillations with a magnitude about 3x larger than those following the original, unshaped pump pulse.
The optimal pulse (c) uses the unsurprising strategy of sending a very short pulse peaked at the relevant exciton transition energies,
$\omega_1 = \SI{12170}{\per\centi\meter}, \omega_2 = \SI{12282}{\per\centi\meter}$.   Although our optimal pulse uses a total duration of $T = \SI{350}{fs}$, all of its meaningful features are in the last \SI{100}{fs}; indeed, our optimizations for different total times between \SI{100}{fs} and \SI{500}{fs}
showed only a small variation in the factor of enhancement of the 1-2 coherence, ranging from 8.4 to 8.6.

To further demonstrate the feasibility and authentication of coherent control of energy transfer in FMO, we now consider using phase-only control to create new states that are not accessible from unshaped (Gaussian) pulses.
This optimization for new states with phase-only control proved to be quite difficult. In most cases, phase-only control only enhances site or exciton populations by at most a few percent in comparison to the unshaped pulse (see Figure~\ref{fig:pop-control}). Verifying such small differences in the population of a specific site or exciton is almost certainly beyond the current experimental state of the art for these open quantum systems \cite{Hoyer:2012ud}.
Achieving phase-only control is especially difficult for natural light harvesting systems because they feature rapid dephasing and energetic relaxation, with each of these operating at a similar time scale to the natural time scale of the energy transfer (see above).  Applying phase-shaping by necessity increases the temporal duration of the pulse, and thus in most cases simply drives the system closer to the equilibrium excited state, rather than to a selected state.

The exact dependence of the pump-probe signal on the density matrix $\rho^{(2)}_\text{pu}(T)$ given by Eq.~\eqref{eq:pump-probe-general} is difficult to determine with high accuracy, even with the assistance of experimental input. Nevertheless, the formal dependence of the signal on the density matrix [Eq.~(\ref{eq:signal})] means that this can be used as a witness for coherently controlled formation of new target states.
Consider the reference set of states through which the system passes following the end of an unshaped Gaussian pump pulse.
Our goal is now to create a state with a spectrum [Eq.~\eqref{eq:signal} with any fixed value of $T$] that falls outside this reference set.
We achieve this by directly optimizing the difference in a simulated pump-probe spectrum from all possible spectra obtained from the reference set of states.
\begin{figure}[]
	\includegraphics[]{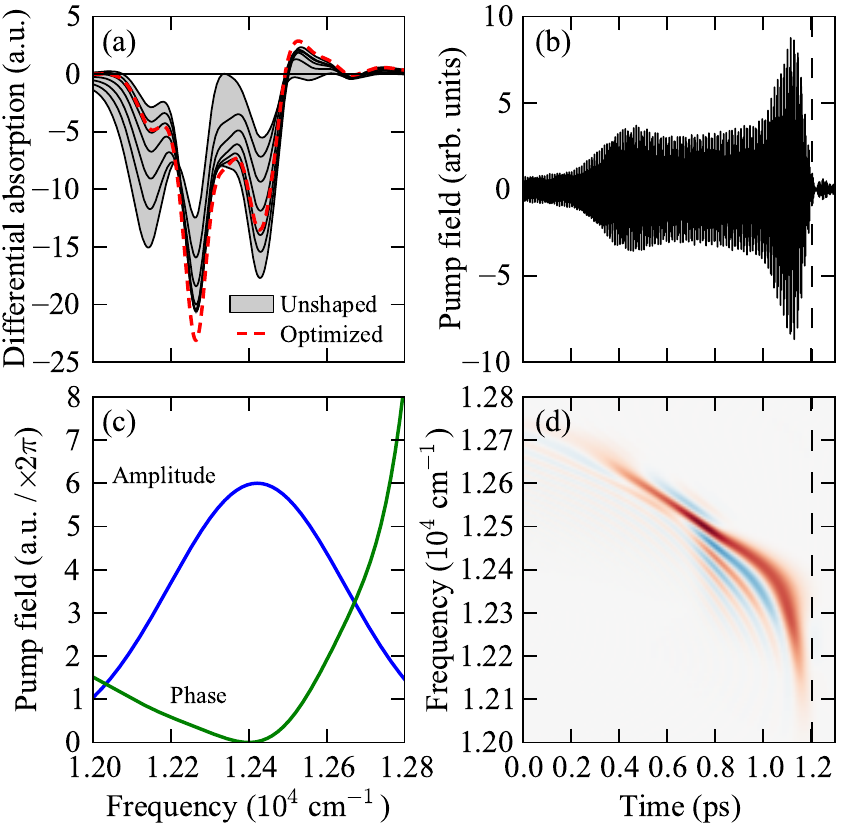}
	\caption{\label{fig:novel-signature}Signature of a novel state in the isotropic pump-probe spectra achieved by phase-only coherent control. The gray area in panel (a) indicates the range of pump-probe spectra (black lines) at various time delays (0, 75, 150, 300, 600 and 1200 fs, and in the limit $T \to \infty$) after the end of the unshaped pump pulse, and the red dashed line shows the spectrum immediately following the optimal pulse. The time-domain, frequency-domain and Wigner spectrogram of the control pulse are shown in panels (b), (c) and (d), respectively. The dashed line in panels (b) and (d) indicates the temporal end of the control pulse. The scale on each plot is in arbitrary units, except for panel (c) which plots the phase in multiples of $2\pi$.}
\end{figure}
With this procedure we found that it is possible to create a 12.0\% relative difference in the signal outside the range of spectra resulting from an unshaped pulse with a time delay. The resulting spectra and visualizations of the optimal pulse constructed from Eq.~\eqref{eq:poly-phase} with $N=10$ terms are shown in Figure \ref{fig:novel-signature}. We chose $N=10$ because our optimizations seemed to reach a point of diminishing returns; in fact, we were able to achieve a maximum difference of 4.2\% with first-order chirp alone ($N=1$) and 9.8\% by adding only one higher order chirp term ($N=2$).

The shapes of the optimized pulses [Figures \ref{fig:pulses}, \ref{fig:coherent-beating}(c) and \ref{fig:novel-signature}(b-d)] are quite complex, as might be expected for coherent control of the subsystem dynamics of this kind of open quantum system in a disordered biological setting. Systematic interpretation of these pulse shapes with the aim of obtaining physical insights for the optimization of coherent dynamics in these systems presents an interesting although challenging goal for future work.  Nevertheless, in many instances, e.g., for the pump-probe spectrum of Figure \ref{fig:novel-signature}(a), some insight can gained by analysis of the time and frequency domain representations of the optimal pulses.  Inspection of the optimal pulse in Figure \ref{fig:novel-signature}(b-c) reveals two distinctive features.  The first is that the optimal pulse sequence is sending in the higher energy light at the earlier times and the second is that there is a distinctive burst of low energy light just before the pulse ends.  We can interpret this as a two-fold route to maximization of excitonic population in low energy states (excitons 2 and 3), since the more energetic excitons initially formed by absorption at the higher frequencies will have time to relax to lower energy states during the delay time. The population in the lower energy states is then further amplified by absorption of a final burst of low energy light just before the pulse ends.

\section{Discussion and outlook}
\label{sec:discussion}

In this paper we have considered the simplest experimental setup for coherent control of excited state dynamics in light harvesting systems, namely a pump-probe configuration.
Most importantly, we showed that in a realistic experimental scenario it should be possible to do two types of coherent control and verification experiments: to create provably new states with phase-only control, and, if both phase and amplitude shaping are available, to enhance quantum beating between excitonic states.

We also analyzed in detail the effect of various restrictions to understand what limits control in light harvesting systems.
The reference closed quantum system given by just the seven chromophores of a single FMO complex is completely controllable, but when the full pigment-protein complex is treated correctly as an ensemble of open quantum systems, the extent of control of the excitonic subsystem is significantly reduced.
Our systematic investigation of the limits of coherent control under the constraints of decoherence at finite temperatures, orientational disorder and static (energetic) disorder,
reveals that each of these three factors restricts the fidelity of achieving both excitonic (energy eigenstates of the electronic Hamiltonian) and site localized target states,
with the fidelity decreasing further when these factors are combined.
Interestingly, our results show that significant improvements in extent of controllability should
however be possible with experiments on well-characterized single molecules. Such single-molecule experiments based on non-linear fluorescence may indeed be possible in the near future \cite{Lott2011, Hildner2013}.

We expect that optimal control will be an even more powerful technique when applied to more complicated non-linear spectroscopy experiments. For example, simultaneous optimization of the first two pulses in a third order photon-echo experiment would allow for preparing particular targeted phase-matched components of the second order density matrix $\rho^{(2)}$. Although
ultimately reflecting the same third order response function, coherent control of the photon echo could be an attractive alternative to full two-dimensional spectroscopy. For higher order spectroscopies the advantages of open loop control for optimal experiment design should be even
greater, given that the exponential growth of the state space of possible experimental configurations makes scanning all possible configurations (as in many 2D experiments) increasingly infeasible.

\begin{acknowledgments}
We thank Jan Roden for assistance with the ZOFE calculations.
This material was supported by DARPA under Award No.\ N66001-09-1-2026.
S.H.\ is a U.S.\ D.O.E.\ Office of Science graduate fellow.
F.C.\ has been supported by the EU FP7 Marie-Curie Programme (Intra-European Fellowship and Career Integration Grant) and by a MIUR-FIRB grant (Project No.\ RBFR10M3SB).
F.C.\ thanks the bwGRiD project and the Imperial College High Performance Computing Service for computational resources.
Sandia National Laboratories is a multi-program laboratory managed and operated by Sandia Corporation, a wholly owned subsidiary of Lockheed Martin Corporation, for the U.S. Department of Energy's National Nuclear Security Administration under contract DE-AC04-94AL85000.
M.B.P.\ is supported by the Alexander von Humboldt Foundation, the ERC Synergy grant BioQ, and the EU Integrating Project Q-ESSENCE and SIQS.
\end{acknowledgments}

\appendix

\section{Orientational average}
\label{sec:orientational-average}

For an ensemble of identical, randomly oriented molecules, it is possible to analytically integrate over all possible molecular orientations with a fixed number of system-field interactions. For two interactions (the case for linear spectroscopies), all signals can be written in the form $\langle S \rangle = \sum_{pq} T_{pq} S_{pq}$, where each sum is over all orthogonal orientations x, y and z and $T_{pq}$ is the second order tensor invariant $\delta_{pq}$. For four interactions (the case for all 3rd order spectroscopies, including pump-probe), all signals can be written in the form $\langle S \rangle = \sum_{pqrs} T_{pqrs} S_{pqrs}$, where each sum is over all orthogonal orientations x, y and z and $T_{pqrs}$ is a linear combination of the fourth order tensor invariants, $\delta_{pq} \delta_{rs}$, $\delta_{pr} \delta_{qs}$ and $\delta_{ps} \delta_{qr}$. The polarization of the pump and probe pulses determines which of these tensor invariants contribute to the observed signal \cite{Hochstrasser2001}. In the Magic Angle configuration, for which the polarization of the first two interactions is at an angle of $54.7\degree$ relative to the polarization of the last two interactions, only the invariant $\delta_{pq} \delta_{rs}$ contributes to the observed signal. Since this invariant is the product of the 2nd order tensor invariant for the first and last two interactions separately, the isotropic averages can be performed separately for the first two (pump) and last two (probe) interactions.

\bibliography{QuControl_FMO}

\end{document}